\documentstyle[12pt]{article}
\def\al{\alpha}

\def\be{\begin{equation}}
\def\ee{\end{equation}}
\def\bea{\begin{eqnarray}}
\def\eea{\end{eqnarray}}

\def\beq{\begin{equation}}
\def\eeq{\end{equation}}
\def\bea{\begin{eqnarray}}
\def\eea{\end{eqnarray}}
\def\nn{\nonumber}
\def\ba{\begin{array}}
\def\ea{\end{array}}   
\def\0{{\mbox{\boldmath $0$}}}
\def\one{1\hskip -.771mm{\rm l}}
\def\A{{\mbox{\boldmath $A$}}}
\def\B{{\mbox{\boldmath $B$}}}

\def\S{{\mbox{\boldmath $S$}}}

\def\vpi{{\mbox{\boldmath $\pi$}}}
\def\r{{\mbox{\boldmath $r$}}}

\def\i{{\rm i}}

\def\ih{\frac{\i}{\hbar}}  
\def\ixh{\i \hbar}

\def\zi{z_{\rm in}}  
\def\zo{z_{\rm out}}

\def\ptwo{\hat{p}_\perp^{\,2}}

\def\pitwo{\hat{\pi}_\perp^{\,2}}

\def\ddz{\frac{\partial}{\partial z}} 
\def\vsig{{\mbox {\boldmath $\sigma$ }}} 
\def\Al{{\mbox{\boldmath $\alpha$}}}

\def\half{\frac{1}{2}}

\def\loh{\lambda_0}

\def\al{\alpha}

\def\E{{\hat{\cal E}}}
\def\O{{\hat{\cal O}}}
\def\eps{\epsilon}
\def\g{\gamma}
\def\Vomeg{{\underline{\mbox {\boldmath $\Omega$}}}_s}

\def\vpip{{\hat{\mbox{\boldmath $\pi$}}}_\perp}

\def\hH{\hat{H}} 
\def\Vsig{{\mbox {\boldmath $\Sigma$ }}} 
\def\Nab{{\mbox{\boldmath $\nabla$}}}

\begin{document}

\begin{flushright}
physics/9809032
\end{flushright}

\begin{center}

{\large\bf Quantum theory of magnetic quadrupole lenses for
spin-$\frac{1}{2}$ particles}\footnote{Presented at the Workshop on 
``Quantum Aspects of Beam Physics'', 15th ICFA (International 
Committee for Future Accelerators) Advanced Beam Dynamics Workshop, 
January 4--9, 1998, Monterey, California, U.S.A.  To appear in the 
Proceedings of the Workshop, Ed. Pisin Chen (World Scientific, 
Singapore, 1998).}

\bigskip

Sameen Ahmed KHAN \\

Dipartimento di Fisica Galileo Galilei  
Universit\`{a} di Padova \\
Istituto Nazionale di Fisica Nucleare~(INFN) Sezione di Padova \\
Via Marzolo 8 Padova 35131 ITALY \\
E-mail: khan@pd.infn.it, ~~~ http://www.pd.infn.it/$\sim$khan/welcome.html

\end{center}

\begin{abstract}
General guidelines for constructing a quantum theory of charged-particle 
beam optics starting {\em ab~initio} from the basic equations of quantum
mechanics, appropriate to the situation under study. In the context of
spin-$\frac{1}{2}$ particles, these guidelines are used starting with the
Dirac equation. The spinor theory just constructed is used to obtain the
transfer maps for normal and skew magnetic quadrupoles respectively. As 
expected the traditional transfer maps get modified by the quantum 
contributions. The classical limit of the quantum formalism presented,
reproduces the well-known Lie algebraic formalism of charged-particle beam
optics. 
\end{abstract}

\bigskip


\bigskip


\newpage

\section{Introduction and Formalism}
Charged-particle beam optics, or the theory of transport of charged-particle
beams  through electromagnetic systems, is traditionally dealt with using 
classical mechanics. This is the case in ion optics, electron microscopy,
accelerator physics etc~\cite{HK}-\cite{Wiedemann} The classical
treatment of charged-particle beam optics has been extremely successful, in
the designing and working of numerous optical devices from electron microscopes
to very large particle accelerators. It is natural to look for a prescription
based on the quantum theory, since any physical system is quantum at the
fundamental level! Such a prescription is sure to explain the grand success of 
the classical theories and may also help towards a deeper understanding and 
designing of certain charged-particle beam devices. To date the curiosity to
justify the success of the classical theories as a limit of a quantum theory 
has been the main motivation to look for a quantum prescription. But, with ever
increasing demand for higher luminosities and the need for polarized beam
accelerators in basic physics, we strongly believe that the quantum theories,
which up till now were an isolated academic curiosity will have a significant
role to play in designing and working of such devices.

It is historically very curious that the, quantum approaches to the
charged-particle beam optics have been very modest and have a very brief
history as pointed out in the third volume of the three-volume encyclopaedic
text book of Hawkes and Kasper~\cite{HK3}.
In the context of accelerator physics the grand success of the classical
theories originates from the fact that 
{\em the de Broglie wavelength of the} ({\em high energy}) 
{\em beam particle is very small compared to the typical apertures of
the cavities in accelerators}. 
This and related details have been pointed out 
in the recent article of Chen~\cite{Chen}.

A beginning of a quantum formalism starting {\em ab initio} with the Dirac
equation was made only recently~\cite{JSSM}-\cite{J2}. The formalism
of Jagannathan{\em et~al} was the first one to use the Dirac equation to
derive the focusing theory of electron lenses, in particular for magnetic
and electrostatic axially symmetric and quadrupole lenses respectively. The
formalism of Jagannathan~{\em et~al} further outlined the recipe to obtain a
quantum theory of aberrations. Details of these and some of the related
developments in {\em the quantum theory of charged-particle beam optics} can
be found in the references~\cite{JSSM}-\cite{JK3}. 
I shall briefly state the central theme of the quantum formalism with some of
the more recent results relevant to accelerator optics.

The starting point to obtain a quantum prescription is to build a theory
based on the basic equations of quantum mechanics appropriate to the situation
under study. For situations when either there is no spin or spinor effects are 
believed to be small and ignorable we start with the scalar Klein-Gordon and
Schr\"{o}dinger equations for relativistic and nonrelativistic cases 
respectively. For electrons, protons and other spin-$\frac{1}{2}$ particles it
is natural to start with the Dirac equation, the equation for
spin-$\frac{1}{2}$ particles. In practice we do not have to care about the 
other~(higher spin) equations.

In many situations the electromagnetic fields are static or can reasonably
assumed to be static. In many such devices one can further ignore the times
of flights which are negligible or of not direct interest as the emphasis is
more on the profiles of the trajectories. The idea is to analyze the
evolution of the beam parameters of the various individual charged-particle
beam optical elements~(quadrupoles, bending magnets,~$\cdots$) along the optic
axis of the system. This in the language of the quantum formalism would 
require to know the evolution of the wavefunction of the beam particles as
a function of `$s$', the coordinate along the optic axis. Irrespective of the
starting basic time-dependent equation~(Schr\"{o}dinger, Klein-Gordon,
Dirac,~$\cdots$) the first step is to obtain an equation of the form
\bea
\i \hbar \frac{\partial }{\partial s} \psi \left(x\, , y\, ; s \right)
=
\hat{\cal H} \left(x\, , y ;\, s \right)
\psi \left(x\, , y\, ; s \right)\,,
\label{BOE}
\eea
where $\left(x , y ; s \right)$ constitute a curvilinear coordinate system,
adapted to the geometry system. For systems with straight optic axis, as it
is customary we shall choose the optic axis to lie along the $Z$-axis and 
consequently we have $s = z$ and
$\left(x , y ; z \right)$ constitutes a rectilinear coordinate system.
Eq.~(\ref{BOE}) is the basic equation in the quantum formalism and we call it
as the {\em beam-optical equation}; ${\cal H}$ and $\psi$ as the 
{\em beam-optical Hamiltonian} and the {\em beam wavefunction} respectively. The
second step requires to obtain a relationship for any relevant
observable $\left\{ \left\langle O \right\rangle \left( s \right) \right\}$ 
at the transverse-plane at $s$ to the observable 
$\left\{ \left\langle O \right\rangle \left( s_{\rm in} \right) \right\}$ 
at the transverse plane at $s _{\rm in}$, where $s _{\rm in}$ is some input
reference point. This is achieved by the integration of the beam-optical
equation in~(\ref{BOE})
\bea
\psi \left(x , y ; s \right) & = &
\hat{U} \left(s , s_{\rm in} \right)
\psi \left(x , y ; s_{\rm in} \right)\,, 
\label{BOI}
\eea
which gives the required transfer maps
\bea
\left\langle O \right\rangle \left( s_{\rm in} \right) 
\longrightarrow
\left\langle O \right\rangle \left( s \right)
& = &
\left\langle \psi \left( x , y ; s \right) 
\left| O \right|
\psi \left( x , y ; s \right) \right\rangle\,, \nn \\
& = &
\left\langle \psi \left( x , y ; s_{\rm in} \right) 
\left| \hat{U} ^{\dagger} O  \hat{U} \right|
\psi \left( x , y ; s_{\rm in} \right) \right\rangle\,.
\label{BOM}
\eea

The two-step algorithm stated above may give an over-simplified picture of 
the quantum formalism than, it actually is. There are several crucial points 
to be noted. The first-step in the algorithm of obtaining the beam-optical 
equation is not to be treated as a mere transformation which eliminates~$t$ 
in preference to a variable~$s$ along the optic axis. A clever set of 
transforms are required which not only eliminate the variable $t$ in preference
to $s$ but also gives us the $s$-dependent equation which has a close physical
and mathematical analogy with the original $t$-dependent equation of standard
time-dependent quantum mechanics. The imposition of this stringent requirement 
on the construction of the beam-optical equation ensures the execution of the
second-step of the algorithm. The beam-optical equation is such, that all the
required rich machinery of quantum mechanics becomes applicable to compute the
transfer maps characterizing the optical system. This describes the essential
scheme of obtaining the quantum formalism. Rest is mostly a mathematical detail
which is built in the powerful algebraic machinery of the algorithm,
accompanied with some reasonable assumptions and approximations dictated by the
physical considerations. For instance, a straight optic axis is a reasonable 
assumption and paraxial approximation constitute a justifiable approximation to
describe the ideal behaviour. 

Before explicitly looking at the execution of the algorithm leading to the
quantum formalism in the spinor case, we further make note of certain other 
features. Step-one of the algorithm is achieved by a set of clever
transformations and an exact expression for the beam-optical Hamiltonian is 
obtained in the case of Schr\"{o}dinger, Klein-Gordon and Dirac equations
respectively, without resorting to any approximations! We expect this to be 
true even in the case of higher-spin equations. The approximations are made
only at step-two of the algorithm, while integrating the beam-optical
equation and computing the transfer maps for averages of the beam parameters.
Existence of approximations in the description of nonlinear behaviour is not 
uncommon and should come as no surprise, afterall the beam optics constitutes
a nonlinear system. The nature of these approximations can be best summarized 
in the optical terminology as; a systematic procedure of expanding the beam
optical Hamiltonian in a power series 
of $\left| {\hat{\vpi} _\perp}/{p_0} \right|$ where $p_0$ is the 
design~(or average) momentum of beam particles moving predominantly along the
direction of the optic axis and $\hat{\vpi} _\perp$ is the small transverse
kinetic momentum. The leading order approximation along with 
$\left| {\hat{\vpi} _\perp}/{p_0} \right| \ll 1$ constitutes the paraxial or 
ideal behaviour and higher order terms in the expansion give rise to the
nonlinear or aberrating behaviour. It is seen that the paraxial and
aberrating behaviour get modified by the quantum contributions which are in
powers of the de Broglie wavelength~($\loh = 2 \pi {\hbar}/{p_0}$). Lastly,
and importantly the question of the classical limit of the quantum formalism;
it reproduces the well known Lie algebraic formalism~\cite{Lie} of 
charged-particle beam optics pioneered by Dragt~{\em et~al}.

Let us start with the Dirac equation for a beam made up of particle of
charge $q$,
mass $m_0$ and anomalous magnetic moment $\mu _a$ in a static electromagnetic
field with potentials $\left(\phi (\r ), \A (\r ) \right)$
\beq
\hat{\rm H}_D \left| \psi_D \right\rangle = E 
\left| \psi_D \right\rangle\,,
\eeq
where $\left| \psi_D \right\rangle$ is the time-independent 
$4$-component Dirac spinor, $E$ is the energy of the beam particle 
and the Hamiltonian $\hat{\rm H}_D$, including the Pauli term is 
\bea 
\hat{\rm H}_D & = & \beta m_0 c^2 + c \Al \cdot (- \ixh \Nab - q \A)
- \mu_a \beta \Vsig \cdot \B \,,  \nn \\
\beta & = &
\left( 
\ba{cc}
\one & \0 \\
\0 & - \one
\ea \right)\,, \quad 
\Al = \left( 
\ba{cc}
\0 & \vsig \\
\vsig & \0 
\ea \right)\,, \quad 
\Vsig = \left( 
\ba{cc}
\vsig & \0 \\
\0 & \vsig 
\ea \right)\,, \nn \\
I & = & \left(
\ba{cc}
\one & \0 \\
\0 &  \one
\ea \right)\,, \quad
\one  =  \left(
\ba{cc}
1 & 0 \\
0 & 1 
\ea \right), \quad
\0 = \left(
\ba{cc}
0 & 0 \\
0 & 0 
\ea \right), \, \nn \\
\sigma_x & = & \left( 
\ba {cc}
0 & 1 \\
1 & 0 
\ea \right), \quad
\sigma_y = \left( 
\ba{cc}
0 & -\i \\
\i & 0 
\ea \right), \quad
\sigma_z = \left( 
\ba{cc}
1 & 0 \\
0 & -1 
\ea \right). \quad   
\eea
We multiply $\hat{\rm H}_D$ (on the left) by $\al _z/c$ and rearrange the 
terms to get 
\bea 
\hat{\cal H}_D & = & - p_0 \beta \chi \al _z - q A_z I + \al _z \Al 
_\perp \cdot \vpip  + (\mu_a /c) \beta \al _z \Vsig \cdot \B\,, \nn \\ 
\chi & = & \left( 
\ba{cc}
\xi \one & \0 \\
\0 & - \xi^{-1} \one
\ea \right)\,, \quad 
\xi = \sqrt{\frac{E + m_0 c^2}{E - m_0 c^2}}\,.
\label{inter}
\eea 
Eq.~(\ref{inter}) is not in the standard form. We need to get
$\hat{\cal H}_D$ into a beam-optical form in close analogy with the
standard Dirac equation, as required by step-one of the algorithm.
Define,
\beq
M  = \frac{1}{\sqrt{2}} (I + \chi \al _z)\,, \qquad \qquad
M^{-1}  = \frac{1}{\sqrt{2}} (I - \chi \al _z)\,,
\eeq
and
\beq
\left| \psi_D \right\rangle 
\longrightarrow
\left| \psi ' \right\rangle 
=
M 
\left| \psi_D \right\rangle\,. 
\eeq
Then
\beq 
\ixh \ddz \left| \psi ' \right\rangle = \hat{\cal H} ' \left| \psi ' 
\right\rangle\,, \quad 
\hat{\cal H} ' =  M \hat{\cal H}_D M^{-1} = - p_0 \beta + \E + \O \,,
\eeq
with the matrix elements of $\E$ and $\O$ given by 
\bea 
\E _{11} & = & - q A_z \one - 
(\mu_a /2c) \left\{ \left(\xi + \xi^{-1} \right) \vsig _\perp \cdot 
\B _\perp + \left( \xi - \xi^{-1} \right) \sigma_z B_z \right\}\,,  
\nn \\
\E _{12} & = & \E _{21} = \0 \,, 
\nn \\
\E _{22} & = & - q A_z \one - 
(\mu_a /2c) \left\{ \left( \xi + \xi^{-1} \right) \vsig _\perp 
\cdot \B _\perp - \left( \xi - \xi^{-1} \right) \sigma_z B_z 
\right\}\,,  
\eea
and
\bea 
\O _{11} & = & \O _{22} = \0 \,, 
\nn \\
\O _{12} & = & \xi\,\left[ \vsig _\perp \cdot \vpip  
- (\mu_a /2c) \left\{ \i \,\left( \xi - \xi^{-1} \right) \left( B_x 
\sigma_y - B_y \sigma_x \right) \right. \right. 
\nn \\ 
  &   & \phantom{- \xi^{-1}\,[ \vsig _\perp \cdot \vpip  
        + (\mu_a /2c) \{ \i} \left. \left.- \left( \xi + \xi^{-1} 
\right) B_z \one \right\} \right]\,, \nn \\
\O _{21} & = & - \xi^{-1}\,\left[ \vsig _\perp \cdot \vpip  
+ (\mu_a /2c) \left\{ \i \,\left( \xi - \xi^{-1} \right) \left( B_x 
\sigma_y - B_y \sigma_x \right) \right. \right. \nn \\
  &   & \phantom{- \xi^{-1}\,[ \vsig _\perp \cdot \vpip 
        + (\mu_a /2c) \{ \i} \left. \left. + \left( \xi + \xi^{-1} 
\right) B_z \one \right\} \right]\,.
\eea
These transformations gives us the beam-optical Hamiltonian
as required by step-one of the algorithm. In the Dirac theory,
the lower spinor components are much smaller than the upper spinor components.
This is also true for the beam-optical equation derived above.
The analogy between the standard Dirac equation and
the beam-optical equation, thus constructed can be summarized in the
following table

\medskip

\begin{tabular}{ll}
{\bf Standard Dirac Equation} ~~~~~~~          & {\bf Beam Optical Form} \\
$m_0 c^2 \beta + \E _D + \O _D$  & $ - p_0 \beta + \E + \O $   \\
$ m_0 c^2 $                          & 
$ - p_0 = \i \hbar \frac{\partial}{\partial z} $ \\
Positive Energy   & Forward Propagation \\
Nonrelativistic, $ \left|\mbox{\boldmath $\pi$}\right| \ll m_0 c$ &
Paraxial Beam, $\left|\mbox{\boldmath $\pi$}_\perp \right| \ll \hbar k$ \\
Non relativistic Motion       & Paraxial Behavior \\
~~ + Relativistic Corrections & ~~ + Aberration Corrections \\
\end{tabular}

\medskip

\noindent
From the above table it is clear that there is a very close analogy between
the derived beam-optical form and the standard Dirac equation. Having
established this analogy we can now execute step-two of the algorithm. As 
is well known the Foldy-Wouthuysen machinery~\cite{FW} enables one to take the
proper nonrelativistic limit of the Dirac equation and provides a systematic
procedure for obtaining the relativistic corrections to a desired degree of
accuracy in powers of ${1}/{m_0 c^2}$. With the analogy in the table above, we
are now in position to adopt the Foldy-Wouthuysen machinery to the beam-optical
situation. The procedure of the Foldy-Wouthuysen-like machinery gives us the
paraxial behaviour and a systematic procedure to compute aberrations to all
orders in powers of ${1}/{p_0}$. To leading order we get
\beq 
\left| \psi ' \right\rangle = \exp \left( \frac{1}{2 p_0}
\beta \O \right) 
\left| \psi^{(1)} \right\rangle\,. 
\eeq
Then
\bea 
\ixh \ddz \left| \psi^{(1)} \right\rangle 
& = & \hat{\cal H}^{(1)} \left| \psi^{(1)} \right\rangle\,, \nn \\ 
\hat{\cal H}^{(1)} & = & \exp \left( - \frac{1}{2 p_0} \beta \O \right) 
\hat{\cal H} ' \exp \left( \frac{1}{2 p_0} \beta \O \right) \nn \\ 
& & \quad - \i \hbar \exp \left(- \frac{1}{2 p_0} \beta \O \right) 
\ddz \left\{ \exp \left(\frac{1}{2 p_0} \beta \O \right) \right\} \nn \\ 
& = & - p_0 \beta + \E ^{(1)} + \O ^{(1)}\,, \nn \\
\E ^{(1)} = \E - \frac{1}{2 p_0} \beta \O ^2 & + & \cdots, \quad
\O ^{(1)} = - \frac{1}{2 p_0} \beta \left\{\left[\O , \E \right] + 
\ixh \ddz \O \right\} + \cdots\,. \quad
\eea 
The lower pair of components of $\left| \psi^{(1)} \right\rangle$ are 
almost vanishing compared to the upper pair and the odd part of 
$\hat{\cal H} ^{(1)}$ is negligible compared to its even part we can 
effectively introduce a Pauli-like two-component spinor formalism based on
the representation.  Calling $\hat{\cal H} ^{(1)}_{11}$ as
$\hat{\tilde{{\cal H}}}$ 
\bea  
\ixh \ddz \left| \tilde{\psi} \right\rangle & = & \hat{\tilde{{\cal{H}}}}
\left| \tilde{\psi} \right\rangle\,, \nn \\
\hat{\tilde{{\cal H}}} 
& \approx & \left( - p_0 - q A_z + \frac{1}{2 p_0} \pitwo \right)  
- \frac{1}{p_0}\left\{ (q + \eps ) B_z S_z + \g \eps 
\B _\perp \cdot \S _\perp \right\}\,, \nn \\
{\rm with ~ } ~ 
\pitwo & = & \hat{\pi}_x^2 + \hat{\pi}_y^2\,, \quad 
\eps = 2 m_0 \mu_a /\hbar\,,  \quad 
\g = E/{m_0 c^2}\,, \quad 
\S = \half \hbar \vsig \,.  
\eea

Up to now, all the observables, the field components, time etc., have been
defined in the laboratory frame. The covariant description of the spin of the
Dirac particle has simple operator representation in the rest frame of the
particle. In the analysis of the spin-dynamics in accelerator physics it is
customary to use such a representation~\cite{Spin}. So we define the following
transform which takes us from the beam-optical form to the
{\em accelerator optical form}
\beq   
\left| \tilde{\psi} \right\rangle = \exp \left\{ \frac{\i}{2 p_0} \left( 
\hat{\pi}_x \sigma_y - \hat{\pi}_y \sigma_x \right) \right\} \left| 
\psi^{(A)} \right\rangle\,. 
\eeq
Details of the {\em accelerator optical transform} are found
elsewhere~\cite{CJKP,Jagan-QABP,CJKP2}. Up to the paraxial approximation
the accelerator-optical Hamiltonian is
\bea 
\i \hbar \ddz \left| \psi^{(A)} \right\rangle & = & \hH ^{(A)} 
\left| \psi^{(A)} \right\rangle\,, 
\nn \\ 
\hH ^{(A)} & \approx & \left( - p_0 - q A_z + \frac{1}{2 p_0}
\pitwo \right) 
+ \frac{\g m_0}{p_0} \Vomeg \cdot \S \,, 
\nn \\ 
{\rm with}\ \ \Vomeg & = & - \frac{1}{\g m_0} \left\{ q \B + 
\eps \left( \B _\parallel + \g \B _\perp \right) \right\}\,.
\eea 
Only at this stage we need to know the specific geometry of the magnetic
lens, i.e., the form of the fields characterizing the lens, for computing
the transfer maps. We shall briefly consider the normal and skew quadrupoles 
respectively in the next sections and see how the traditional transfer maps
get modified by the quantum contributions.

{\section{Example-I: Normal Magnetic Quadrupole}
An ideal normal magnetic quadrupole of length $\ell$ is characterized 
by the field
\beq 
\B = (-Gy , -Gx , 0), 
\eeq
corresponding to the vector potential
\beq 
\A = \left( 0 , 0 , \half G \left( x^2 - y^2 \right) \right)\,, 
\eeq 
situated in the transverse-planes $z = \zi$ and $z = \zo + \ell$, with 
$G$ constant inside the lens and zero outside. The 
accelerator-optical Hamiltonian is
\beq 
\hH (z) = \left\{ 
\ba{l}
\hH _F = - p_0 + \frac{1}{2 p_0} \ptwo \,, 
\quad {\rm for}\ \ z < \zi \ \ {\rm and}\ \ z > \zo \,,\\ 
\hH _L(z) = - p_0 + \frac{1}{2 p_0} \ptwo - \half q G \left( x^2 - y^2 \right) 
+ \frac{\eta p_0}{\ell} \left( y \sigma_x + x \sigma_y \right)\,, \\ 
\qquad \qquad {\rm for}\ \ \zi \leq z \leq \zo \,, \quad 
{\rm with}\ \ \eta = (q + \g \eps )G\ell\hbar/{2 p_0 ^2}\,.  
\ea \right. 
\eeq   
The subscripts $F$ and $L$ indicate, the field-free and the lens region 
respectively.

Best way to compute $\hat{U}$ is {\em via} the interaction picture, 
used in the Lie algebraic formulation~\cite{Lie} of 
classical beam optics. Using the transfer operator thus
derived~(details found elsewhere~\cite{CJKP,CJKP2}) we get the
transfer maps for averages with the subscripts {\em in} and {\em out}
standing for $(z_{\rm in})$ and $(z_{\rm out})$ respectively
\bea 
& & \left( 
\ba{c}
\langle x \rangle \\  \\ 
{\langle \hat{p}_x \rangle } /p_0 \\   \\
\langle y \rangle \\  \\ 
{\langle \hat{p}_y \rangle}/{p_0} \\
\ea \right) _{\rm out}
\approx 
T_{\rm q}
\left( \left( 
\ba{c}            
\langle x \rangle \\   \\ 
{\langle \hat{p}_x \rangle}/{p_0} \\  \\
\langle y \rangle \\  \\ 
{\langle \hat{p}_y \rangle}/{p_0} \\
\ea \right) 
+ \eta \left( 
\ba{c}
\left(\frac{\cosh\,\left(\sqrt{K}\,\ell \right) - 1}{K \ell} \right) 
\langle \sigma _y \rangle \\ 
- \left( \frac{\sinh\,\left(\sqrt{K}\,\ell \right)} {\sqrt{K}\,\ell}\right) 
\langle \sigma _y \rangle \\ 
- \left( \frac{\cos\,\left(\sqrt{K}\,\ell \right) - 1}{K \ell} \right) 
\langle \sigma _x \rangle \\ 
- \left( \frac{\sin\,\left(\sqrt{K}\,\ell \right)} {\sqrt{K}\,\ell}\right) 
\langle \sigma _x \rangle 
\ea \right) \right)_{\rm in}\,, \nn \\
& & \nn \\
& &
M_{\rm q} =
\left( 
\ba{cccc}
\cosh (\sqrt{K} \ell) 
& \frac{1}{\sqrt{K}}\sinh (\sqrt{K} \ell) 
& 0 & 0 \\ 
\sqrt{K} \sinh (\sqrt{K} \ell) 
& \cosh (\sqrt{K} \ell)
& 0 & 0 \\ 
0 & 0 &
\cos (\sqrt{K} \ell) 
& \frac{1}{\sqrt{K}}\sin (\sqrt{K} \ell) \\   
0 & 0 &
- \sqrt{K} \sin (\sqrt{K} \ell) 
& \cos (\sqrt{K} \ell)
\ea \right)\,, \nn \\ 
& & 
M_{\stackrel{<}{>}} =
\left(
\ba{cccc}
1 & \Delta z_{\stackrel{<}{>}} & 0 & 0  \\
0 & 1 & 0 & 0 \\
0 & 0 & 1 & \Delta z_{\stackrel{<}{>}} \\
0 & 0 & 0 & 1 \\
\ea \right), \qquad \qquad 
T_{\rm q} = M_> M_{\rm q} M_<\,.
\eea
For spin, the transfer map reads
\bea
\langle S_x \rangle_{\rm out}
&\approx&
\langle S_x \rangle_{\rm in} \nn \\
& & + \frac{4 \pi \eta}{\loh} 
\left(\left(\frac{\sinh (\sqrt{K} \ell)}{\sqrt{K} \ell}\right)
\langle x S_z \rangle_{\rm in} 
+ \left(\frac{\cosh (\sqrt{K} \ell) - 1}{K \ell p_0}\right) 
\langle \hat{p} _x S_z \rangle_{\rm in} \right)\,, \nn \\ 
\langle S_y \rangle_{\rm out} 
& \approx &
\langle S_y \rangle_{\rm in} \nn \\ 
& & - \frac{4 \pi \eta}{\loh}
\left(\left(\frac{\sin (\sqrt{K} \ell)}{\sqrt{K} \ell}\right) 
\langle y S_z \rangle_{\rm in}
- \left(\frac{\cos (\sqrt{K}\,\ell) - 1}{K \ell p_0}\right) 
\langle \hat{p}_y S_z \rangle_{\rm in}\right)\,, \nn \\ 
\langle S_z \rangle_{\rm out} 
& \approx &
\langle S_z \rangle_{\rm in} \nn \\ 
& & - \frac{4 \pi \eta}{\loh}
\left\{\left(\frac{\sinh (\sqrt{K} \ell)}{\sqrt{K} \ell}\right) 
\langle x S_x \rangle_{\rm in}
- \left(\frac{\sin (\sqrt{K} \ell)}
{\sqrt{K} \ell}\right)
\langle y S_y \rangle_{\rm in} \right. \nn \\
& & \quad \left.
+ \left(\frac{\cosh (\sqrt{K} \ell) - 1}{K \ell p_0}\right)
\langle \hat{p}_x S_x \rangle_{\rm in}
+ \left(\frac{\cos (\sqrt{K} \ell) - 1}{K \ell p_0}\right)
\langle \hat{p}_y S_y \rangle_{\rm in}
\right\}\,.
\eea 
Thus, we get the a fully quantum mechanical derivation for the traditional 
transfer maps~(transfer matrices)~\cite{Wiedemann}. In addition we also get
the spinor contributions, the Stern-Gerlach kicks as they are called. In
recent years there has been a campaign to make a 
spin-splitter~\cite{Splitter}
device to produce polarized beams using the Stern-Gerlach kicks. The spinor
theory of charged-particle beam optics, {\em in principle} supports the
spin-splitter devices. 

\section{Example-II: Skew Magnetic Quadrupole} 
For a skew-magnetic-quadrupole lens the field is given by
\beq
\B = \left(- G_s y , G_s x , 0 \right),
\eeq
corresponding to the vector potential
\beq
\A = \left( 0 , 0 , - G_s x y \right).
\eeq
The accelerator-optical Hamiltonian is 
\beq
\hH (z) = \left\{ 
\ba{l}
\hH _F = - p_0 + \frac{1}{2 p_0} \ptwo \,, 
\quad {\rm for}\ \ z < \zi \ \ {\rm and}\ \ z > \zo \,,\\ 
\hH _L(z) = - p_0 + \frac{1}{2 p_0} \ptwo + \half q G_s x y 
- \frac{\eta _s p_0}{\ell} \left( x \sigma_y - y \sigma_x \right)\,, \\ 
\quad \quad 
{\rm for}\ \ \zi \leq z \leq \zo \,, \quad 
{\rm with}\ \ \eta _s = (q + \g \eps )G_s \ell\hbar/{2 p_0 ^2}\,.  
\ea \right. 
\eeq
and the corresponding transfer operator is
\bea
& & \hat{\tilde{U}}_{i,L} (\zo , \zi) \nn \\
& & = \exp \left\{- \ih \frac{\eta _s}{2}
\left[\left(
- \frac{S^{+}}{\sqrt{K_s} \ell} p_0 x
+ \frac{C^{-}}{K_s \ell} \hat{p}_x
+ \frac{S^{-}}{\sqrt{K_s} \ell} p_0 y
+ \frac{\left(C^{+} - 2 \right)}{K_s \ell}
\hat{p}_y \right) \sigma_y \right. \right. \nn \\
& & \qquad \qquad \left. \left. - \left(
- \frac{S^{-}}{\sqrt{K_s} \ell} p_0 x
+ \frac{\left(C^{+} - 2 \right)}{K_s \ell} \hat{p}_x
- \frac{S^{+}}{\sqrt{K_s} \ell} p_0 y
+ \frac{C^{-}}{K_s \ell}
\hat{p}_y \right) \sigma_x \right]\right\}\,, \nn \\ 
& & 
C^{\pm} = \cos (\sqrt{K_s} \ell)
\pm \cosh (\sqrt{K_s} \ell), 
\quad
S^{\pm} = \sin (\sqrt{K_s} \ell)
\pm \sinh (\sqrt{K_s} \ell)\,.
\eea
Then we get the transfer maps
%
\bea
& &  
\left( 
\ba{c}
\langle x \rangle \\ \\ 
{\langle \hat{p}_x \rangle}/{p_0} \\ \\
\langle y \rangle \\ \\ 
{\langle \hat{p}_y \rangle}/{p_0} \\ \\
\ea \right)_{\rm out}
\approx 
T_{\rm sq}
\left( \left( 
\ba{c}
\langle x \rangle \\ \\
{\langle \hat{p}_x \rangle}/{p_0} \\ \\
\langle y \rangle \\ \\
{\langle \hat{p}_y \rangle}/{p_0} \\ \\
\ea \right) 
+ \frac{\eta _s}{2}
\left( 
\ba{c}
\frac{- C^{-} \langle \sigma_y \rangle 
      + \left(C^{+} - 2 \right) \langle \sigma_x \rangle}{{K_s} \ell} \\  \\
\frac{-S^{+} \langle \sigma_y \rangle 
      + S^{-} \langle \sigma_x \rangle}{\sqrt{K_s} \ell}\\  \\
\frac{- \left(C^{+} -2 \right) \langle \sigma_y \rangle
      + C^{-} \langle \sigma_x \rangle}{{K_s} \ell} \\ \\
\frac{- S^{-} \langle \sigma_y \rangle 
       + S^{+} \langle \sigma_x \rangle}{\sqrt{K_s} \ell} \\
\ea \right) \right)_{\rm in}\,, \nn \\
& & 
M_{\rm sq} = \frac{1}{2}
\left(
\ba{cccc}
C^{+} & \frac{S^{+}}{\sqrt{K_s}} & C^{-} & \frac{S^{-}}{\sqrt{K_s}} \\
- \sqrt{K_s} S^{-} & C^{+} & - \sqrt{K_s} S^{+} & C^{-} \\
C^{-} & \frac{S^{-}}{\sqrt{K_s}} & C^{+} & \frac{S^{+}}{\sqrt{K_s}} \\
- \sqrt{K_s} S^{+} & C^{-} & - \sqrt{K_s} S^{-} & C^{+} \\
\ea \right), 
\quad
T_{\rm sq} = M_> M_{\rm sq} M_<\,.
\eea
Again we get a quantum derivation of the traditional transfer 
maps~\cite{Wiedemann} with some spinor contributions. Due to the algebraic
machinery of the quantum formalism it is straightforward to see that the
transfer maps of the skew quadrupole can be obtained by rotation of the 
transfer maps for the normal quadrupole by ${\pi}/4$ along the optic axis. 
For spin, the transfer map reads
\bea
\langle S_x \rangle _{\rm out}
& \approx & \langle S_x \rangle _{\rm in}
+ \frac{2 \pi \eta _s}{\loh} 
\left\{- \left(\frac{S^{+}}{\sqrt{K_s} \ell}\right) 
\langle x S_z \rangle _{\rm in} 
+ \left(\frac{C^{-}}{K_s \ell p_0}\right) 
\langle \hat{p} _x S_z \rangle _{\rm in} \right. \nn \\ 
& & \quad \qquad \qquad \qquad \left. 
- \left(\frac{S^{-}}{\sqrt{K_s} \ell}\right) 
\langle y S_z \rangle _{\rm in} 
+ 
\left(\frac{C^{+} - 2}{K_s \ell p_0}\right) 
\langle \hat{p}_y S_z \rangle _{\rm in}\right\}\,, \nn \\ 
\langle S_y \rangle _{\rm out} 
& \approx & \langle S_y \rangle _{\rm in} 
+ \frac{2 \pi \eta _s}{\loh} 
\left\{- \left(\frac{S^{-}}{\sqrt{K_s} \ell}\right) 
\langle x S_z \rangle _{\rm in}
 + 
\left(\frac{C^{+} -2}{K_s \ell p_0}\right) 
\langle \hat{p}_x S_z \rangle _{\rm in} \right. \nn \\ 
& & \quad \qquad \qquad \qquad \left. 
- \left(\frac{S^{+}}{\sqrt{K_s}\,\ell}\right) 
\langle y S_z \rangle _{\rm in}
+ 
\left(\frac{C^{-}}{K_s \ell p_0}\right) 
\langle \hat{p}_y S_z \rangle _{\rm in} \right\}\,, \nn \\ 
 & & \nn \\
\langle S_z \rangle _{\rm out}
& \approx & \langle S_z \rangle _{\rm in}          
+ \frac{2 \pi \eta _s}{\loh} \left\{ \left\{
\left(\frac{S^{+}}{\sqrt{K_s} \ell}\right) 
\langle x S_x \rangle _{\rm in}
- \left(\frac{C^{-}}{K_s \ell p_0}\right) 
\langle \hat{p}_x S_x \rangle _{\rm in} \right. \right. \nn \\ 
& & \quad \qquad \qquad \qquad \left. 
+ \left(\frac{S^{-}}{\sqrt{K_s} \ell}\right)
\langle y S_x \rangle _{\rm in}
- \left(\frac{C^{+} - 2}{K_s \ell p_0}\right) 
\langle \hat{p}_y S_x \rangle _{\rm in} \right\} \nn \\ 
& & \quad \qquad \qquad \quad
+
\left\{- \left(\frac{S^{-}}{\sqrt{K_s} \ell}\right) 
\langle x S_y \rangle_{\rm in}
+ \left(\frac{ C^{+} -2}{K_s \ell p_0}\right)
\langle \hat{p}_x S_y \rangle_{\rm in} \right. \nn \\ 
& & \quad \qquad \qquad \qquad \left. \left. 
- \left(\frac{S^{+}}{\sqrt{K_s} \ell}\right) 
\langle y S_y \rangle_{\rm in}
+ \left(\frac{C^{-}}{K_s \ell p_0}\right) 
\langle\hat{p}_y S_y \rangle_{\rm in}
\right\}\right\}.
\eea 

\section{Concluding Remarks and Directions for Future Research}
To summarize, an algorithm is presented for constructing a 
{\em quantum theory of charged-particle beam optics}, starting
{\em ab~initio} from the basic equations of quantum mechanics appropriate to
the situation. This algorithm is used to construct a spinor theory of
accelerator optics starting {\em ab~initio} from the Dirac equation, taking 
into account the anomalous magnetic moment. The formalism is demonstrated by
working out the examples of the normal and skew magnetic quadrupoles 
respectively and as expected there are small quantum contributions, which are 
proportional to the powers of the de Broglie wavelength. It is
further shown that the quantum formalism presented, in the classical limit
reproduces the Lie algebraic formalism~\cite{Lie} of charged-particle 
beam optics.

The present algorithm is suited for constructing a quantum theory of
charged-particle beam optics at the single-particle level. The next logical
step would be to extend it to accommodate the realistic multiparticle
dynamics. We feel that clues to such a formalism can come from the experience 
gained from the so-called quantum-{\em like} models~\cite{Fedele}.
These phenomenological models have been extensively developed in recent years 
to explain the classical collective behavior of a charged-particle beam, by
associating with the classical beam a quantum-{\em like} wavefunction obeying a 
Schr\"{o}dinger-{\em like} equation with the role of $\hbar$ being played by 
the beam emittance $\varepsilon$.  

One practical application of the quantum formalism would be to get a deeper
understanding of the polarized beams. A proposal to produce polarized beams
using the proposed spin-splitter devices based on the classical Stern-Gerlach
kicks has been presented recently~\cite{Splitter}.

Lastly it is speculated that the quantum theory of charged-particle beam
optics will be able to resolve the {\em choice of the position 
operator} in the Dirac theory and the related question of the
{\em form of the force
experienced by a charged-particle in external electromagnetic 
fields}~\cite{Heinemann}.
This will be possible provided one can do an extremely high precision 
experiment to detect the small differences arising in the transfer maps from
the different choices of the position operators. These differences shall be 
very small, i.e., proportional to powers of the de Broglie wavelength. It is
the extremely small magnitude of these minute differences which makes the
exercise so challenging and speculative!

\section*{Acknowledgments}
In this First Book on {\it Quantum Aspects of Beam Physics} I would like to
thank Prof. R. Jagannathan, for all my training in the exciting field of
{\em quantum theory of charged-particle beam optics}. It is a pleasure to thank 
Prof. Pisin Chen, and the Organizing Committee of QABP98, for my participation 
in the historic Monterey meeting.
I am thankful Prof. M. Pusterla for kind encouragement and my thanks
are due to him, and to INFN-Padova, for providing full financial support for my
travel to participate in QABP98.

\end{document}